\documentclass[prd,preprint,tightenlines,floatfix,
preprintnumbers,nofootinbib,eqsecnum]{revtex4}
\usepackage[dvips,final]{graphicx}
\usepackage{amssymb,amsmath,epsfig,bm,pifont}
\usepackage{pstricks,pst-node,pst-text,pst-3d}
\usepackage{caption}
\usepackage{multirow}
\usepackage{tabularx}
\usepackage{makecell}
\usepackage{floatrow}
\def\1{\hbox{{1}\kern-.25em\hbox{l}}}

\begin{document} 

\begin{flushright}
INR-TH-2018-034
\end{flushright}

\title{The least squares method: application to analysis of the  flavor dependence of the QCD relation between pole and $\rm{\overline{MS}}$-scheme running heavy quark masses}
\vspace*{0.8cm}

\author{A. L. Kataev and V. S. Molokoedov}
{\let\thefootnote\relax\footnote{e-mail: kataev@ms2.inr.ac.ru\\ e-mail:  viktor\_molokoedov@mail.ru}}

\address{Institute for Nuclear
Research RAS, 117312 Moscow, Russia\\~\\
Moscow Institute of Physics and Technology, 141700 Dolgoprudny, Russia}

\date{\today}

\begin{abstract}
The features of the ordinary least squares method, which gives a possible way to a solution of the overdetermined systems of algebraic equations and allows to estimate the uncertainties of the obtained solutions, are considered. As the important physical example we define four-loop QCD coefficients in the dependence of the relation between  pole and  running heavy quarks masses  on the number of light flavors, using the existing results of numerical supercomputer based calculations of the corresponding four-loop contributions at different fixed numbers of light flavors. Stability of the found solutions to the number of the considered equations and unknowns is demonstrated and supported by the Pearsons's $\chi$-squared test.
\end{abstract}

\maketitle
\thispagestyle{empty}

\newpage
\setcounter{page}{1}

\section{Introduction}

The problem of the asymptotic character of the series of perturbation theory (PT) in quantum field theory has been attracted the attention of the theoreticians for quite a long time (see e.g. \cite{Basuev:1974yd} and references therein). In the $\phi^4$-theory definite progress in its 
description was made in the well-known work \cite{Lipatov:1976ny}. The  similar approach, though with less predictive ability, was applied for analysis of the asymptotic structure of the perturbative QED series in  
Ref.\cite{Itzykson:1977mf} and of the Yang-Mills theory in Ref.\cite{Bogomolny:1977ty} (for review of these and other 
related works see e.g.\cite{Kazakov:1980rd}). 

In the studies of the asymptotic nature of the perturbative QCD series for the physical quantities, evaluated in the $\rm{\overline{MS}}$-scheme, the renormalon-based methods are more widely used 
(see e.g.\cite{Zakharov:1997xs}, \cite{Beneke:1998ui, Beneke:2000kc}, \cite{Kataev:2005hv}).  In this work we consider the QCD series of PT for relation between pole and $\rm{\overline{MS}}$-scheme  running masses of heavy quarks 
with its infrared renormalon (IRR) asymptotic structure, derived in Refs.\cite{Bigi:1994em} and  \cite{Beneke:1994sw}. 

This relation, which sometimes briefly called as the $\rm{\overline{MS}}$-on-shell relation, read
\begin{equation}
\label{expansion}
z_m(\mu^2)=\frac{\overline{m}_q(\mu^2)}{M_q}=1+\sum\limits_{i=1}^{\infty} z^{(i)}_m a^i_s(\mu^2),
\end{equation}
where $\overline{m}_q(\mu^2)$ and $M_q$ are the $\rm{\overline{MS}}$-scheme running scale-dependent and pole (or on-shell (OS)) masses of heavy quarks correspondingly, $a_s=\alpha_s/\pi$, $\alpha_s$  is the QCD coupling constant, defined in the $\rm{\overline{MS}}$-scheme. 

The one-, two- and three-loop terms $z^{(1)}_m, z^{(2)}_m, z^{(3)}_m$ were calculated in Refs.\cite{Tarrach:1980up}, \cite{Gray:1990yh, Avdeev:1997sz, Fleischer:1998dw} and \cite{Melnikov:2000qh, Chetyrkin:1999qi} respectively. In the case of the color gauge group $SU(3)$ these coefficients, normalized by the condition $\mu^2=M^2_q$, have the following numerical form
\begin{eqnarray}
\label{z1-3}
z^{(1)}_m=-4/3, ~~~ z^{(2)}_m=-14.332+1.0414n_l, ~~~ 
z^{(3)}_m=-198.71+26.924n_l-0.6527n^2_l,
\end{eqnarray}    
where corrections of the second and third orders of PT depend on the number of massless quarks $n_l$. Note that we consider here the case of one heavy flavor and $n_l$ massless ones, i.e. the number of active quarks $n_f=n_l+1$. Moreover, it is easy to understand from the diagrams, which are responsible for renormalization of the two-point Green function of quarks fields, that the contribution $z^{(i)}_m$ of the $i$-th order of PT is the polynomial of $(i-1)$-th degree in $n_l$. Indeed, in the $i$-th order of PT the gluon propagator, renormalizing scalar two-point quark correlator, contains $(i-1)$ inserts of fermion loops,  each of which gives a factor, proportional to $n_l$.  In the presented below studies we focus on the four-loop contribution $z^{(4)}_m$ of  Eq.(\ref{expansion}) and its cubic polynomial expansion in powers of $n_l$. The numerical values of the $z^{(4)}_m$-term for $n_l=3,4,5$  were first obtained at the Super-Computer Lomonosov of MSU \cite{Lomonosov} with the help of semi-analytical calculations in Ref.\cite{Marquard:2015qpa} and later found with higher  precision at various fixed values of $0\leq n_l\leq 20$  in  Ref.\cite{Marquard:2016dcn}.  

At the  first glance it may seem that the four-loop contribution in the relation between pole and running masses in QCD is not huge. However, the situation is not so transparent. Indeed, as already mentioned above,  the  PT  QCD  series  for these physical quantities are asymptotic with the IRR dominated  factorially growing coefficients. Therefore,  it is important from both phenomenological and theoretical points of view to fix the number of order of PT from which the asymptotic nature of the PT series will manifest itself. This task is more relevant for the cases of bottom and top-quark pole masses.  Indeed, for $c$-quark the  $\mathcal{O}(a^3_s)$-contribution  to the $\rm{\overline{MS}}$-on-shell relation has already exceeded the $\mathcal{O}(a^2_s)$-term, and in the case of the $b$-quark three-loop correction is still remains less than two-loop one.
Hence, to understand better when the asymptotic structure of the 
relation between pole and running masses of the $b$ (and heavier $t$) quark is manifesting itself it is important to know the magnitude of the $\mathcal{O}(a^4_s)$-correction and its $n_l$-dependence.

In this work we study the problem of the flavor dependence with the help of the application of the ordinary least squares method (LSM), previously 
used in Refs.\cite{Kataev:2015gvt, Kataev:2018gle}. It was formulated  by Gauss and independently by Legendre  long time ago ($\sim$ in  1800) to find   approximate solutions of the overdetermined systems of algebraic equations.
 This method  has basically two main applications: the first, the most widespread and well-known, permits to determine the uncertainties of the unknown parameters, entering into the set of the solved equations, while 
  the second, less commonly used at present, allows to evaluate the central values of the solutions of the overdetermined systems of these equations  (see e.g.\cite{Linnik}).

\section{Application of the least squares method}

\subsection{The case of four unknowns}

As  was already noted above the expression for $z^{(4)}_m$-contribution in Eq.(\ref{expansion}) is written as 
\begin{equation}
\label{z4-m}
z^{(4)}_m=z^{(40)}_m+z^{(41)}_mn_l+z^{(42)}_mn^2_l+z^{(43)}_mn^3_l.
\end{equation}
To get the values of four coefficients $z^{(4s)}_m$ at $0\leq s\leq 3$ in Eq.(\ref{z4-m}) we apply the LSM to the results of numerical calculations \cite{Marquard:2016dcn} of the four-loop $z^{(4)}_m$-contribution with the 
corresponding $n_l$-dependent mean-square uncertainties
at fixed number of light flavors from the wide range $0\leq n_l\leq 20$. However, in our analysis we restrict ourselves by the consideration of the results of Ref.\cite{Marquard:2016dcn} from the interval  $3\leq n_l\leq 15$, where the lower bound corresponds to the real number of existing  heavy quarks, while the upper bound matches to the number of  massless flavors at which the QCD property of the asymptotic freedom does not violated $(n_l< 31/2)$. Indeed, this condition follows from the negativity of the first coefficient $\beta_0$ of the QCD $\beta$-function, viz $\beta_0=-1/4(11-2/3(n_l+1))< 0$ \cite{Gross:1973id, Politzer:1973fx}. Taking into account the foregoing we derive the following overdetermined system of equations for the coefficients $z^{(4s)}_m$ normalized at $\mu^2=M^2_q$:
\begin{equation}
\label{system}
\begin{pmatrix}
    1 & 3 & 9 & 27 \\ 1 & 4 & 16 & 64 \\ 1 & 5 & 25 & 125 \\ 1 & 6 & 36 & 216 \\ 1 & 7 & 49 & 343 \\ 1 & 8 & 64 & 512 \\ 1 & 9 & 81 & 729 \\ 1 & 10 & 100 & 1000 \\ 1 & 11 & 121 & 1331 \\ 1 & 12 & 144 & 1728 \\ 1 & 13 & 169 & 2197 \\ 1 & 14 & 196 & 2744 \\ 1 & 15 & 225 & 3375 \\ 
   \end{pmatrix}
   \begin{pmatrix}
   z^{(40)}_m \\ 
   \\  z^{(41)}_m \\
   \\ z^{(42)}_m \\
   \\ z^{(43)}_m \\   
   \end{pmatrix}
   = \begin{pmatrix} 
   -1756.36\pm 1.74 \\
   -1278.70\pm 1.77 \\
   -871.73\pm 1.80 \\
   -531.39\pm 1.84 \\
   -253.59\pm 1.87\\
   -34.28\pm 1.91\\
   130.62\pm 1.94\\
   245.17\pm 1.98\\
   313.45\pm 2.01\\
   339.51\pm 2.05\\
   327.44\pm 2.08\\
   281.30\pm 2.12\\
   205.16\pm 2.16\\
   \end{pmatrix}
   \end{equation}
In accordance with LSM we have to enter the $\Psi$-function, which is  
equal to sum of the squares of deviations of all equations in the system (\ref{system}):
\begin{equation}
\label{Psi-4}
\Psi(z^{(40)}_m, z^{(41)}_m, z^{(42)}_m, z^{(43)}_m)=\sum\limits_{k=1}^{N} (z^{(40)}_m+z^{(41)}_m n_{l_k}+z^{(42)}_mn^2_{l_k}+z^{(43)}_mn^3_{l_k}     -f_{l_k})^2~,
\end{equation}
where index $k$ takes all values equal to the number $N$ of equations in the system (\ref{system}) (in this particular case we have $N=13$), 
$f_{l_k}$ is the column of the  numbers in r.h.s. of this system with their uncertainties $\Delta f_{l_k}$.

The LSM solutions of the overdetermined system (\ref{system}) correspond to  the such values of $z^{(4s)}_m$-parameters, for which the $\Psi$-function has a minimum, defined by the following requirements\footnote{Note, that conditions (\ref{conditions}) basically coincide with the requirements, postulated in the method of minimal sensitivity 
\cite{Stevenson:1981vj}, which is actively used now in the studies related to the scheme-dependent ambiguities of the massless PT QCD series for 
renormalization-group invariant quantities.}:
\begin{equation}
\label{conditions}
\frac{\partial\Psi}{\partial z^{(40)}_m}=0, ~~~ \frac{\partial \Psi}{\partial z^{(41)}_m}=0, ~~~ \frac{\partial\Psi}{\partial z^{(42)}_m}=0, ~~~ \frac{\partial\Psi}{\partial z^{(43)}_m}=0.
\end{equation}
These conditions lead to the system of 4 equations of the form $G_{is}z^{(4s)}_m=F_i$ with 4 unknowns $z^{(40)}_m, z^{(41)}_m, z^{(42)}_m, z^{(43)}_m$, where matrix $G$ is the Gram matrix. As is known the Gram matrix is a symmetric positive-definite matrix and therefore the solution of the system, obtained from the conditions (\ref{conditions}), exists and is unique. For $N=13$ we have ${\rm{det}}(G)=97538785344$.

After solution of the system (\ref{conditions}) we can fix LSM uncertainties of the obtained by us values of $z^{(4s)}_m$-terms using the law of accumulation of errors:
\begin{subequations}
{\small
\begin{gather}
\label{z4-a}
\Delta z^{(40)}_m=\sqrt{\sum\limits_{k=1}^{N} \left( \frac{\partial z^{(40)}_m}{\partial f_{l_k}} \Delta  f_{l_k} \right)^2}=\frac{1}{{\rm{det}}(G)}\sqrt{\sum\limits_{k=1}^{N} \Delta  f^2_{l_k}\begin{pmatrix}{\rm{det}}\begin{pmatrix}
1 & c_1 & c_2 & c_3 \\
n_{l_k} & c_2 & c_3 & c_4 \\
n^2_{l_k} & c_3 & c_4 & c_5 \\
n^3_{l_k} & c_4 & c_5 & c_6 \\
\end{pmatrix}
\end{pmatrix} ^2}~, \\
  \label{z4-b}
\Delta z^{(41)}_m=\sqrt{\sum\limits_{k=1}^{N} \left( \frac{\partial z^{(41)}_m}{\partial f_{l_k}} \Delta  f_{l_k} \right)^2}=\frac{1}{{\rm{det}}(G)}\sqrt{\sum\limits_{k=1}^{N} \Delta  f^2_{l_k}\begin{pmatrix}{\rm{det}}\begin{pmatrix}
c_0 & 1 & c_2 & c_3 \\
c_1 & n_{l_k} & c_3 & c_4 \\
c_2 & n^2_{l_k} & c_4 & c_5 \\
c_3 & n^3_{l_k} & c_5 & c_6 \\
\end{pmatrix}
  \end{pmatrix} ^2}~, \\
  \label{z4-c}
  \Delta z^{(42)}_m=\sqrt{\sum\limits_{k=1}^{N} \left( \frac{\partial z^{(42)}_m}{\partial f_{l_k}} \Delta  f_{l_k} \right)^2}=\frac{1}{{\rm{det}}(G)}\sqrt{\sum\limits_{k=1}^{N} \Delta  f^2_{l_k}\begin{pmatrix}{\rm{det}}\begin{pmatrix}
c_0 & c_1 &  1 & c_3 \\
c_1 & c_2 &  n_{l_k} & c_4 \\
c_2 & c_3 &  n^2_{l_k} & c_5 \\
c_3 & c_4 &  n^3_{l_k} & c_6 \\
\end{pmatrix}
  \end{pmatrix} ^2}~, \\
  \label{z4-d}
   \Delta z^{(43)}_m=\sqrt{\sum\limits_{k=1}^{N} \left( \frac{\partial z^{(43)}_m}{\partial f_{l_k}} \Delta  f_{l_k} \right)^2}=\frac{1}{{\rm{det}}(G)}\sqrt{\sum\limits_{k=1}^{N} \Delta  f^2_{l_k}\begin{pmatrix}{\rm{det}}\begin{pmatrix}
c_0 & c_1 & c_2 &  1 \\
c_1 & c_2 & c_3 &  n_{l_k} \\
c_2 & c_3 & c_4 &  n^2_{l_k} \\
c_3 & c_4 & c_5 &  n^3_{l_k} \\
\end{pmatrix}
  \end{pmatrix} ^2}~,
\end{gather}}
\end{subequations}
where $c_p=\sum\limits_{j=1}^N n^p_{l_j}$. The combined use of formulas (\ref{conditions}) and (\ref{z4-a}-\ref{z4-d}) leads to the following results:
\begin{eqnarray}
\label{answ-401}
&&z^{(40)}_m=-3654.16\pm 7.21, ~~~~~~ z^{(41)}_m=756.95\pm 2.98, \\ \label{answ-423}
&&z^{(42)}_m=-43.48\pm 0.37, ~~~~~~~~~ z^{(43)}_m=0.678\pm 0.014.
\end{eqnarray}
It should be emphasized  that the obtained expressions of $z^{(4s)}_m$-terms demonstrate the sign-alternating in $n_l$ structure of the four-loop contribution $z^{(4)}_m$, which is observed at the  two- and three-loop levels of PT QCD results as well. 

\subsection{The case of two unknowns}

In fact the coefficients of the  leading and sub-leading in $n_l$ terms in expansion (\ref{z4-m}) are known in analytical form from the calculations of  Ref.\cite{Lee:2013sx}. The obtained  numerical expression  of $z^{(43)}_m$-coefficient is in agreement with its value previously found in Ref.\cite{Ball:1995ni} in the process of evaluating the contribution from the renormalon-chain of fermion one-loop insertions into the
 $\rm{\overline{MS}}$-on-shell heavy quark mass relation.
 These exactly computed coefficients have the following numerical form: 
\begin{equation} 
\label{exact z43-42}
z^{(43)}_m=0.67814, ~~~~~~ z^{(42)}_m=-43.4824.
\end{equation}
It should be stressed  that the results (\ref{exact z43-42}) are in agreement with the central values of the corresponding terms in Eq.(\ref{answ-423}), obtained by means of the LSM.

It is interesting now to study whether the LSM-expressions (\ref{answ-401}) will be affected  by fixing the explicitly known numbers of Eq.(\ref{exact z43-42}). Combining them with the results of calculations \cite{Marquard:2016dcn} we transform the system of Eqs.(\ref{system})
to the similar one, which contains two unknown coefficients $z^{(40)}_m$ and $z^{(41)}_m$ only:
\begin{equation}
\label{system-1}
\begin{pmatrix}
    1 & 3 \\ 1 & 4 \\ 1 & 5  \\ 1 & 6  \\ 1 & 7 \\ 1 & 8 \\ 1 & 9 \\ 1 & 10 \\ 1 & 11 \\ 1 & 12 \\ 1 & 13 \\ 1 & 14 \\ 1 & 15 \\ 
   \end{pmatrix}
   \begin{pmatrix}
   z^{(40)}_m \\ \\
   \\  z^{(41)}_m \\
   \end{pmatrix}
   = \begin{pmatrix} 
   -1383.33\pm 1.74 \\
   -626.38\pm 1.77 \\
   130.56\pm 1.80 \\
   887.50\pm 1.84 \\
   1644.45\pm 1.87\\
   2401.39\pm 1.91\\
   3158.33\pm 1.94\\
   3915.27\pm 1.98\\
   4672.22\pm 2.01\\
   5429.15\pm 2.05\\
   6186.09\pm 2.08\\
   6943.03\pm 2.12\\
   7699.98\pm 2.16\\
   \end{pmatrix}
   \end{equation}
In this case the analogs of formulas (\ref{Psi-4}) and (\ref{conditions}) take the following form:
\begin{equation}
\label{Psi-2}
\Psi(z^{(40)}_m, z^{(41)}_m)=\sum\limits_{k=1}^{N} (z^{(40)}_m+z^{(41)}_m n_{l_k}-y_{l_k})^2 ~~~~ \text{and} ~~~~\frac{\partial\Psi}{\partial z^{(40)}_m}=0, ~~ \frac{\partial \Psi}{\partial z^{(41)}_m}=0,
\end{equation}
where $y_{l_k}$ are the  numbers, presented on the r.h.s. of the system (\ref{system-1}) with  $N=13$   linear  equations. The expressions for the corresponding  LSM-uncertainties have the following simplified  representation:
{\small
\begin{gather}
\nonumber
\Delta z^{(40)}_m=\sqrt{\sum\limits_{k=1}^{N} \left( \frac{\partial z^{(40)}_m}{\partial y_{l_k}} \Delta  y_{l_k} \right)^2}=
\frac{1}{N\sum\limits_{k=1}^{N} n^2_{l_k}-\bigg(\sum\limits_{k=1}^{N} n_{l_k}\bigg)^2}\sqrt{\sum\limits_{k=1}^{N} \Delta y^2_{l_k}\bigg(\sum\limits_{i=1}^{N} n^2_{l_i}-n_{l_k}\sum\limits_{i=1}^{N} n_{l_i}\bigg)^2}, \\ 
\label{unc}
\Delta z^{(41)}_m=\sqrt{\sum\limits_{k=1}^{N} \left( \frac{\partial z^{(41)}_m}{\partial y_{l_k}} \Delta  y_{l_k} \right)^2}=\frac{1}{N\sum\limits_{k=1}^{N} n^2_{l_k}-\bigg(\sum\limits_{k=1}^{N} n_{l_k}\bigg)^2}\sqrt{\sum\limits_{k=1}^{N} \Delta y^2_{l_k}\bigg(N \;n_{l_k}-\sum\limits_{i=1}^{N} n_{l_i}\bigg)^2}.
\end{gather}}
Applying formulas (\ref{Psi-2}) and (\ref{unc}) we obtain the numerical values for the constant and linearly dependent on $n_l$ terms to the four-loop contribution $z^{(4)}_m$ with their theoretical inaccuracies:
\begin{equation}
\label{res}
z^{(40)}_m=-3654.14\pm 1.34, ~~~~~ z^{(41)}_m=756.94\pm 0.15.
\end{equation}
As can be seen from results (\ref{answ-401}) and (\ref{res}) the central values of $z^{(40)}_m$ and $z^{(41)}_m$-terms practically do not change when the number of unknowns is halved. At the same time the uncertainties of these coefficients are reduced noticeably. 

The task of solving the system (\ref{system-1}) has a simple geometric interpretation: it is necessary to draw a straight line optimally based on the given 13 points on the plane in the coordinates $y_l(n_l)$. In this case coefficients $z^{(40)}_m$ and $z^{(41)}_m$ define the angles of inclination of the line with the axes $n_l$ and $y_l$. Therefore there is nothing strange in that the uncertainties of the obtained coefficients are less than the ones, presented in column in the r.h.s. of system (\ref{system}). Similarly, it is not a weird  that inaccuracies in  Eqs.(\ref{answ-423})  exceed the ones of Eq.(\ref{res}).

The values (\ref{res}) should be compared with the results, derived in Ref.\cite{Marquard:2016dcn}:
\begin{equation}
\label{prev}
z^{(40)}_m=-3654.15\pm 1.64, ~~~~~ z^{(41)}_m=756.942\pm 0.040.
\end{equation}
Despite the fact that the central values of the results (\ref{res}) and (\ref{prev}) are obtained within the various approaches, they coincide.  It should be stressed, that the result (\ref{prev}) of $z^{(40)}_m$-term  was fixed in \cite{Marquard:2016dcn} as the value of the four-loop contribution  $z^{(4)}_m$ at  $n_l=0$ and it did not take into account  the correlation effects with other expressions, obtained at nonzero values of $n_l$, whereas the result (\ref{res}) is extracted from the data for  
 $3\leq n_l\leq 15$   and therefore takes into consideration these effects. As the result the uncertainty of $z^{(40)}_m$-term in Eq.(\ref{res}) is sligly smaller than one, presented in Ref.\cite{Marquard:2016dcn}. 

The  interesting effects are observed when the physical number of heavy quarks flavors are  considered \cite{Kataev:2015gvt} only, namely $3\leq n_l \leq 5$. In this case the LSM system read
\begin{equation}
\label{system-2}
\begin{pmatrix}
    1 & 3 \\ 1 & 4 \\ 1 & 5  \\  
   \end{pmatrix}
   \begin{pmatrix}
   z^{(40)}_m \\ 
   \\  z^{(41)}_m \\
   \end{pmatrix}
   = \begin{pmatrix} 
   -1383.33\pm 1.74 \\
   -626.38\pm 1.77 \\
   130.56\pm 1.80 \\ 
   \end{pmatrix}.
   \end{equation}
Application of Eqs.(\ref{Psi-2}) and (\ref{unc}) leads to the next result:
\begin{equation}
\label{point-3}
z^{(40)}_m=-3654.16\pm 5.08, ~~~~~ z^{(41)}_m=756.95\pm 1.25.
\end{equation}
The central values of Eqs.(\ref{point-3}) are almost indistinguishable from the ones, given in Eqs.(\ref{answ-401}) and (\ref{res}).
Therefore, we conclude that the LSM results are stable to the number of equations in the considered overdetermined systems.

The observed stability of the central values, obtained within the LSM, can mainly be explained by the fact that quantities $\xi_k=y_{l_k}-z^{(40)}_m-z^{(41)}_mn_{l_k}$ (see Eq.(\ref{Psi-2})) form a sample of values of a normal random quantity with mathematical expectation close to zero and some variance $\sigma^2$. This statement can be verified using the Pearson's $\chi$-squared test (see e.g.\cite{Sachs}). Of course, for a reliable answer to this question it is necessary to have a sufficiently large number of input sample points, that is not observed in the considered cases $(N=13)$. However, as a first approximation we can estimate the value of the Pearson's $\chi^2$-parameter for our problem. For this aim we should build a grouped statistical series of absolute frequencies. First of all we determine $\xi_{min}=min\{\xi_k\}$, $\xi_{max}=max\{\xi_k\}$, fix the number $m$ of grouping intervals and the length of these intervals $h=(\xi_{max}-\xi_{min})/m$. Secondly, we find the right bounds of group intervals $\hat{\xi}_j=\xi_{min}+jh$ and their centers $\tilde{\xi}_j=\hat{\xi}_j+h/2$, where $1\leq j \leq m$. The absolute frequencies $\tilde{n}_j$ are  defined as the number of elements $\xi_k$ belonging to the interval $(\hat{\xi}_{j-1}; \hat{\xi}_j)$. Taking this discussion into account one can obtain the following Table:

\begin{table}[!h]
\begin{center}
\caption{Table of the absolute and theoretical frequencies}
{\def\arraystretch{1.5}\tabcolsep=0.1pt
\begin{tabular}{|c|c|c|c|c|}
\hline 
$~~~j~~~$ & ~~~~$\hat{\xi}_j$~~~~  & ~~~~~$\tilde{\xi}_j$~~~~~ & ~~~$\tilde{n}_j$~~~ & ~~~$n_j$~~~ \\
\hline 
1 & $\;$ -0.004 $\;$ &  -0.001  & 1  & 1.569 \\
\hline
2 & 0.002 & 0.005 & 3 & 3.399 \\
\hline
3 & 0.008 & 0.011 & 6 & 5.003  \\
\hline
4 & 0.014 & 0.017 & 2 & 2.912 \\
\hline
5 & 0.020 & 0.023 & 1 & 0.670  \\
\hline
\end{tabular}}
\end{center}
\end{table}

In this Table we use the following data $\xi_{min}=-0.01$, $\xi_{max}=0.02$, $m=5$, $h=0.006$. The mathematical expectation is equal to $\overline{\xi}=\sum\limits_{j=1}^m \tilde{\xi}_j\tilde{n}_j/N=0.0105$, which is close to zero, whereas the variance is equal to $\sigma^2=\sum\limits_{j=1}^m \tilde{n}_j(\tilde{\xi}_j-\overline{\xi})^2/(N-1)=0.0000388$.

At the next stage we should fix the theoretical frequencies $n_j$, obtained for the normal distribution law, which are the product of the total number of random quantities $\xi_k$ and the probability $p_j=h\Phi((\tilde{\xi}_j-\overline{\xi})/\sigma)/\sigma $, where $\Phi$-function is the standard normal distribution function $\Phi(u)=\exp(-u^2/2)/\sqrt{2\pi}$. The theoretical non-rounded frequencies $n_j$, calculated in this way, are represented in the Table.

Now everything is ready for application of the $\chi$-squared test. For this goal we consider the value $\chi^2=\sum\limits_{j=1}^m (\tilde{n}_j-n_j)^2/n_j=0.9$. In our case the number of statistical degrees of freedom
 is equal to $v=m-s-1=2$, where parameter $s$ denotes 2 degrees of freedom of the normal distribution law. At the standard level of the significance $\alpha=0.05$ according to the table of critical distribution points $\chi^2$ we find that $\chi^2_{crit}(v=2; \alpha=0.05)=5.99$. Thus we conclude that $\chi^2=0.9<\chi^2_{crit}=5.99$ and this means that the hypothesis about the normal distribution of random variables $\xi_k$ is confirmed.

Therefore the maximum likelihood function will have a form close to Gaussian, namely $L=(2\pi\sigma^2)^{-N/2}\exp\bigg(-\sum\limits_{k=1}^N (\xi_k-\overline{\xi})^2/(2\sigma^2)\bigg)$. Since the mathematical expectation is really small in comparison with the contributions $y_{l_k}$, presented on the r.h.s. of the system (\ref{system-1}), then the function $L$ achieves the maximal value at those values of $z^{(40)}_m$ and $z^{(41)}_m$-terms when the conditions (\ref{Psi-2}) are held. Thus, now the meaning of the LSM requirements (\ref{Psi-2}) becomes clear from the point of view of statistical mathematics.

All above mentioned discussions reflect the elegance of the LSM, which is a truly powerful fitting procedure. Moreover, as we have seen the LSM allows to check the self-consistency of the results of the four-loop  numerical computations, presented in \cite{Marquard:2016dcn}, and  outcomes of  analytical four-loop calculations, performed in Ref.\cite{Lee:2013sx}. The numerical values of the fourth order contributions to the relation between 
pole and $\rm{\overline{MS}}$-scheme running masses of charm, bottom and top-quarks with taking into account the LSM-results for $z^{(40)}_m$ and $z^{(41)}_m$-terms can be found in Refs.\cite{Kataev:2015gvt, Kataev:2018gle}.

\section{Conclusion}

Applying the ordinary method of the least squares to the overdetermined system of algebraic equations with $3\leq n_l\leq 15$ we define not only the values of the two yet unknown in analytical form coefficients in the four-loop contribution to relation between pole and running masses of heavy quarks but also fix their corresponding uncertainties. The central values of these terms are consistent with a high degree of accuracy with the results
of the numerical calculations, presented in Ref.\cite{Marquard:2016dcn}. To demonstrate the stability of the least squares method to the number of equations and the number of unknowns variables we consider two separate situations: when the number of equation is equal to 3, namely $3\leq n_l\leq 5$, and when we do not take into account the results of analytical computations for the leading $n^3_l$ and sub-leading $n^2_l$-terms \cite{Lee:2013sx}. It is interesting to note that in  both cases the central values of all unknown terms are almost the same as the previously obtained values at $3\leq n_l \leq 15$, while them uncertainties increase no more than 10 times. The validity of application of the least squares method is explained by the Pearson's $\chi$-squared test and the maximum likelihood method. The presented in this work description of the LSM clarifies the special features of its applications for the determination of the 
explicit dependence on the number of lighter flavors of the four-loop approximations between pole and $\rm{\overline{MS}}$-scheme running heavy quarks masses. Naturally, the application of the LSM is not limited only to this problem and it also may be used in future for solving other tasks, where it is necessary to obtain the generalized solutions of the overdetermined systems of algebraic equations.

\end{document}